\documentstyle[aps,preprint,prl,epsfig]{revtex} 

\draft
\newcommand{\ba}{\begin{eqnarray}}
\newcommand{\ea}{\end{eqnarray}}
\newcommand{\bmath}{\begin{mathletters}}
\newcommand{\emath}{\end{mathletters}}
\newcommand{\ban}{\begin{eqnarray*}}
\newcommand{\ean}{\end{eqnarray*}}
\newcommand{\tl}{\tilde{\ell}}
\begin{document}

\title{Consequences of a Relativistic Pseudospin Symmetry\\ 
for Radial Nodes and Intruder Levels in Nuclei}
\author{A. Leviatan$^{1,*}$ and J.N. Ginocchio$^{2,\dagger}$}
\address{$^{1}$~Racah Institute of Physics, The Hebrew University,
Jerusalem 91904, Israel}
\address{$^{2}$~Theoretical Division, Los Alamos National Laboratory, Los
Alamos, New Mexico 87545, USA}
\date{\today}

\maketitle

\begin{abstract}
The identification of pseudospin symmetry as a relativistic symmetry 
of the Dirac Hamiltonian is used to explain the structure of 
radial nodes occurring in pseudospin doublets and to illuminate 
the special status of nodeless intruder states in nuclei. 
\end {abstract}

\vspace{4cm}
\noindent $^{*}$E-mail address: ami@vms.huji.ac.il\\
$^{\dagger}$E-mail address: gino@t5.lanl.gov\\

\newpage

The concept of pseudospin symmetry \cite{aa,kth} 
is based on the empirical observation 
of quasi-degenerate pairs of certain normal-parity
shell-model orbitals with non-relativistic quantum numbers
\ba
\left (\,n_{r},\ \ell,\ j = \ell + {1\over 2}\,\right )\;\;\;
{\rm and}\;\;\;
\left (\,n_{r}-1,\ \ell + 2,\ j = \ell + {3\over 2}\,\right ) ~.
\label{psdoub}
\ea
Here $n_r$, $\ell$, and $j$ are the
single-nucleon radial, orbital, and total angular momentum quantum
numbers, respectively. The doublet structure, 
is expressed in terms of a ``pseudo'' orbital angular momentum
$\tilde{\ell}$ = $\ell$ + 1 and ``pseudo'' spin, $\tilde s$ = 1/2, which 
are coupled to total angular momentum $j = \tilde{\ell}\pm \tilde s$. 
For example, $(n_r s_{1/2},(n_r-1) d_{3/2})$ will have
$\tilde{\ell}= 1$,
$(n_r p_{3/2},(n_r-1) f_{5/2})$ will have $\tilde{\ell}= 2$, etc.
This pseudospin symmetry has been used to explain 
features of deformed nuclei \cite{bohr}, 
including superdeformation \cite{dudek}
and identical bands \cite{twin,stephens} and to establish 
an effective shell-model coupling scheme \cite{trolt}. In view 
of its central role in both spherical and deformed nuclei, there has been 
an intense effort to understand the origin of this symmetry. 
The observed reduction of pseudo spin-orbit splitting in the non-relativistic 
single-particle spectra was shown to follow from nuclear relativistic 
mean-fields \cite{bahri,draayer}. The pseudospin symmetry itself has 
been shown to arise from a relativistic symmetry of a Dirac Hamiltonian 
in which the sum of the scalar, $V_S$, and vector, $V_V$, potentials 
cancel, $V_S + V_V =0$ \cite{gino,ami}. 
An attractive scalar and repulsive vector potentials of
nearly equal magnitudes, $V_S \sim - V_V$, is an inherent feature of 
realistic relativistic mean fields in nuclei. Calculations 
employing such fields in a variety of nuclei have confirmed the 
existence of an approximate pseudospin symmetry in the energy spectra 
and wave functions \cite{gino2,ring,arima,ginoami}. 
The conditions for an approximate relativistic pseudospin 
symmetry have been further elaborated \cite{marcos} including the effect 
of isospin asymmetry \cite{alberto}. 
Implications on magnetic properties and Gamow-Teller transitions 
in nuclei \cite{m1gt} and on nucleon-nucleus scattering have been 
considered \cite{scat}. 

In the relativistic pseudospin scheme \cite{gino,ami}, 
the non-relativistic wave functions
of Eq.~(\ref{psdoub}) are associated with the upper components of 
Dirac wave functions which are eigenstates of a Dirac Hamiltonian with 
$V_S\sim -V_V$. 
The pseudospin $\tilde{s}$, and pseudo orbital angular momentum $\tl$, 
are found to be the ordinary spin, and ordinary orbital angular momentum 
respectively, of the lower component of the Dirac wave functions. 
The underlying Dirac structure ensures that the wave function of the 
upper component of the Dirac eigenfunction has a spherical harmonic of 
rank $\ell= \tl-1$ for aligned spin: $j=\tl-1/2=\ell+1/2$, and a 
spherical harmonic of rank $\ell+2=\tl+1$ for unaligned spin: 
$j=\tl +1/2=(\ell+2) - 1/2$. This explains the particular angular 
momenta defining the pseudospin doublet of Eq.~(\ref{psdoub}). 
A prominent and uncommon feature of this doublet is that its members 
have different radial wave functions. 
In particular, the state with 
aligned spin in Eq.~(\ref{psdoub}) has $n_r$ nodes as compared to  
$n_r-1$ nodes of its partner with unaligned spin. One objective of 
the present letter is to reveal the so far unexplained 
mechanism by which the relativistic 
origin of pseudospin symmetry ensures this particular relation 
between radial quantum numbers of pseudospin partners. 
A related question concerns the shell-model states with aligned spin and 
no nodes, {\it e.g.}, $0s_{1/2},\; 0p_{3/2},\; 0d_{5/2}$, etc. 
For heavy nuclei such states with large $j$, {\it i.e.}, 
$0g_{9/2},\;0h_{11/2},\;0i_{13/2}$, are the ``intruder'' 
abnormal-parity states which, due to the strong spin-orbit term, 
intrude from the shell above and defect to the shell below. 
The original formulation of pseudospin was oriented to a symmetry of 
normal-parity shell-model orbitals only. The intruder states 
were discarded from the non-relativistic pseudospin scheme. 
However, that is clearly not a satisfactory procedure if pseudospin is 
identified as a relativistic symmetry of the Dirac Hamiltonian, since 
now both the normal-parity states and abnormal-parity states are eigenstates 
of the same Hamiltonian. The relativistic attributes of pseudospin 
symmetry will allow us to examine the properties and special status of 
these intruder levels and to explain why these states do not have a partner 
which is an eigenstate.

Properties of Dirac wave functions in a central field were considered 
in \cite{rose1,rose2} for vector potentials only. The discussion below 
generalizes these results to the case where both scalar, $V_S(r)$,  
and vector, $V_V(r)$, spherically symmetric potentials are present in 
the Dirac Hamiltonian. As usual, in this case the eigenstates 
can be written in the form $\Psi_{\kappa,m} = \left ( 
g_{\kappa}\left [Y_{\ell}\,\chi\right]^{(j)}_{m}\,,\,
if_{\kappa}\left [Y_{\ell'}\,\chi\right]^{(j)}_{m}\,\right )$ 
where $g_{\kappa}(r)$ and $f_{\kappa}(r)$ 
are the radial wave functions of the upper and lower components 
respectively, $Y_{\ell}$ and $\chi$ are the spherical harmonic and 
spin function which are coupled to angular momentum $j$ projection $m$.
The labels $\kappa = -(j+1/2)$ and $\ell'= \ell +1$ for aligned spin 
$j=\ell+1/2$, while $\kappa = (j+1/2)$ and $\ell'= \ell -1$ 
for unaligned spin $j=\ell-1/2$. Setting $\hbar=c=1$, the 
radial Dirac equations for a single nucleon of mass $M$ and total 
energy $E$ are
\bmath
\ba
{dG\over dr} &=& -{\kappa\over r}G + 
A(r)\,F \\
{dF\over dr} &=& {\kappa\over r}F - 
B(r)\,G
\ea
\label{dirac1}
\emath
where $G=rg_{\kappa}$, $F=rf_{\kappa}$ and 
\bmath
\ba
A(r) &=& \left [\, E + M + V_S(r) - V_V(r) \, \right ] \\
B(r) &=& \left [\, E - M - V_S(r) - V_V(r) \, \right ] ~. 
\ea
\label{ab}
\emath
We consider potentials satisfying 
$rV_S,\; rV_V \rightarrow 0$ for $r\rightarrow 0$, 
and $V_S,\; V_V \rightarrow 0$ for $r\rightarrow \infty$. 
For bound state solutions $-M < E < M$ and the radial wave functions 
satisfy $G(0) = F(0) =0$, $G(\infty) = F(\infty) = 0$, 
and $\int_{0}^{\infty} ( G^2 + F^2 )\, dr = 1$.
Specifically, we are interested in bound Dirac valence states for which
both the binding energy $(M-E) > 0 $ and the total energy $E>0$ are 
positive. For relativistic mean fields relevant to nuclei, 
$V_S$ is attractive and $V_V$ is repulsive with typical values 
$V_S(0) \sim -400,\; V_V(0)\sim 350,\; E\sim 900,\; M \sim 940$ MeV. 
Accordingly, in practical applications, the quantity $A(r)$ in Eq.~(3a) is 
positive definite, $A(r) > 0$, with $A(0)\sim 1090$ MeV, while 
the quantity $B(r)$ in Eq.~(3b) is monotonic in $r$ and changes sign 
from a small positive value at the origin, $B(0)\sim 10$ MeV, 
to a negative value $(E-M)<0$ at large $r$. 
It is now straightforward to obtain from Eq.~(\ref{dirac1}) the asymptotic 
behavior of $G$ and $F$,  
\ba
\begin{array}{lllll}
{\rm for}\; r\longrightarrow \infty \; &
G \sim e^{-\lambda r} \;,\; & F \sim e^{-\lambda r} \;,\; &  
F/G \longrightarrow -\sqrt{{M-E\over M+E}} \; , \;\;
\lambda = \sqrt{M^2-E^2}\\
{\rm for\; r\longrightarrow 0} \; &  
G \sim r^{-\kappa}\;,\; & F \sim r^{1-\kappa} \; , \; &
F/G \longrightarrow 
-{B(0)\over 1 -2\kappa}\,r \, < 0 \; , \;\;\;
(\kappa < 0) \\
{\rm for\; r\longrightarrow 0} \; & 
G \sim r^{1+\kappa}\;,\; & F \sim r^{\kappa} \; , \; & 
G/F \longrightarrow \;\; 
{A(0)\over 1 +2\kappa}\,r \, > 0 \; , \;\;\;\,
(\kappa > 0) ~.
\end{array}
\label{rsmallrlarge}
\ea
Both $G$ and $F$ behave as a power law for small $r$ and exhibit an 
exponential falloff for large $r$.  
We illustrate this behavior of $G$ and $F$ in Fig.~1a,c,e for the 
$2p_{3/2}$ $(\kappa=-2)$, $1f_{5/2}$ $(\kappa=3)$ and 
$0g_{9/2}$ $(\kappa=-5)$ eigenstates of a Dirac Hamiltonian with scalar 
and vector potentials of Woods-Saxon form with parameters tuned to the 
neutron spectra of $^{208}$Pb \cite{alberto}.

To study further properties of the radial wave functions, 
it is convenient to introduce 
${\cal G} = r^{\kappa}G$ and ${\cal F} = r^{-\kappa}F$.
Then in the open interval $(0,\infty)$, 
nodes of $F$ and $G$ coincide with nodes of ${\cal F}$ and ${\cal G}$.
From Eq.~(\ref{dirac1}) get 
\bmath
\ba
{d{\cal G}\over dr} &=& 
r^{2\kappa}\, A(r)\,{\cal F} \\
{d{\cal F}\over dr} &=& 
-r^{-2\kappa}\, B(r)\, {\cal G} ~.
\ea
\label{dirac2}
\emath
A number of observations follow from Eq.~(\ref{dirac2}). 
First, we note that it is impossible 
for ${\cal F}$ and ${\cal G}$, or $F$ and $G$, to vanish 
simultaneously at the same point because if they did, then 
all other higher-order derivatives would vanish at that point and hence 
the functions themselves would vanish everywhere. 
Second, we see that 
a node of ${\cal F}$ corresponds to an extremum of ${\cal G}$, and a node 
of ${\cal G}$ corresponds to an extremum of ${\cal F}$ 
(since $B(r)$ in Eq.~(5b) changes sign, ${\cal F}$ can have an additional 
extremum where $B(r)=0$, which does not correspond to a node of ${\cal G}$, 
but this can occur only at one point since $B(r)$ is monotonic). 
It follows that the nodes of $F$ and $G$ alternate; that is, between 
every pair of adjacent nodes of $F$ (or $G$) there is one node of $G$ 
(or $F$). If we let $r_1$ be a node of ${\cal F}$ and $r_2$ be a node of 
${\cal G}$, then the nature of the extrema at these points is determined 
from the second derivatives
\bmath
\ba
{d^{2}{\cal G}\over dr^{2}}\bigg\vert_{r=r_1} &=&
-A(r_1)B(r_1)\,{\cal G}(r_1) \quad 
{\rm where\;\;} {\cal F}(r_1) = 0 \\
{d^{2}{\cal F}\over dr^{2}}\bigg\vert_{r=r_2} &=&
-A(r_2)B(r_2)\,{\cal F}(r_2) \quad 
{\rm where\;\;} {\cal G}(r_2) = 0 ~.
\ea
\label{2der}
\emath
As bound states, both ${\cal G}$ and ${\cal F}$ vanish at 
$r=\infty$ and their extrema are concave towards the $r$-axis. Therefore,  
the extrema at the nodes $r_1$ or $r_2$ are minima (maxima) 
if the functions ${\cal G}$ or ${\cal F}$ are negative (positive) 
respectively at these points. It follows from Eq.~(\ref{2der}) 
that nodes of ${\cal F}$ and ${\cal G}$ can occur only where 
the product $A(r)B(r) > 0 $ is positive. Since for practical applications 
$A(r)>0$, this condition reduces to 
\ba
B(r) \; > 0 && \;\;
{\rm at \; nodes\; of\; {\cal F}\; and \; {\cal G}}~.
\label{condpos}
\ea
The combination $V_S + V_V$ appearing in $B(r)$, Eq.~(3b), 
is the average potential felt by the nucleon in the non-relativistic 
limit. Therefore, the condition of Eq.~(\ref{condpos}) is similar to the 
statement in the non-relativistic case, that nodes of the 
radial wave function can occur only in the region of classically allowed 
motion, {\it i.e.}, where the kinetic energy is positive. 

We now use the above results to obtain a relation between the radial nodes 
of $F$ and $G$. For that purpose we consider the equation 
for $GF={\cal G}{\cal F}$ as derived from Eq.~(\ref{dirac1}), 
\ba
\left ( GF \right )' &=& A(r)\, F^2 - B(r)\, G^2 ~.
\label{GFprime}
\ea
For large $r$, 
$\left ( GF \right )' \sim (E+M)F^2 - (E-M)G^2 \, > 0$ 
is positive, since the binding energy $(M-E) > 0$ for bound states. 
At small $r$, 
$\left ( GF \right )' = -B(0)\,G^2 < 0$ for $\kappa < 0$, while 
$\left ( GF \right )' = A(0)\,F^2 > 0$ for $\kappa > 0$, 
by employing Eq.~(\ref{rsmallrlarge}). 
Since $GF$ vanishes both at $r=0$ and $r=\infty$ we see that it is 
an increasing negative function at large $r$, while at small $r$, 
$GF$ is a decreasing negative function for $\kappa<0$ and 
an increasing positive function for $\kappa>0$,  
\bmath
\ba
& {\rm for}\; r\longrightarrow \infty \qquad &  GF < 0 \quad  \\
& {\rm for\; r\longrightarrow 0} \qquad &  GF < 0 \quad (\kappa < 0) \\
& {\rm for\; r\longrightarrow 0} \qquad &  GF > 0 \quad  (\kappa > 0)
\ea
\label{gfsign}
\emath
consistent with Eq.~(\ref{rsmallrlarge}). 
Furthermore, since $A(r) > 0$ and by using Eq.~(\ref{condpos}) 
we find that 
\ba
\begin{array}{ll}
\left ( GF \right )'\bigg\vert_{r=r_1} =  
-B(r_1) \, G^{2}(r_1) \; < 0 \quad
& {\rm where\;\;} F(r_1) = 0 \\
\left ( GF \right )'\bigg\vert_{r=r_2} = 
\;
A(r_2)\, F^{2}(r_2) \; > 0 \quad
& {\rm where\;\;} G(r_2) = 0 ~. \\
\end{array}
\label{gfzero}
\ea
Thus $GF$ is a decreasing function at the nodes of $F$, and an increasing 
function at the nodes of $G$. 
Exploiting all these derived properties, we observe that for $\kappa > 0$, 
$GF$ is positive at small $r$ and negative at large $r$, and hence has
an odd number of zeroes. By Eq.~(\ref{gfzero}) the first and last zeroes 
of $GF$ correspond to nodes of $F$, and 
since the nodes of $F$ and $G$ alternate, then 
the number of nodes of $F$ exceed by one the number of nodes 
of $G$. On the other hand, for $\kappa < 0$, $GF$ has the same 
(negative) sign near both end points, and hence has an even number of 
zeroes. By similar arguments we find that in this case the first and 
last zeroes of $GF$ are nodes of $G$ and $F$ respectively, and 
that $G$ and $F$ have the same number of nodes. These properties of $GF$ 
are illustrated in Fig.~1b,d for the $2p_{3/2}$ $(\kappa=-2)$, 
$1f_{5/2}$ $(\kappa=3)$ states. Altogether we have,
\ba
\begin{array}{ll}
\kappa < 0: \quad & n_F = n_G \qquad \\ 
\kappa > 0: \quad & n_F = n_G + 1 
\end{array}
\label{nodes}
\ea
where $n_F$ and $n_G$ denote the number of internal nodes of $F$ and $G$ 
respectively. For $\kappa > 0$, the first and last nodes (considering 
$F$ and $G$ together) are $F$ nodes. For $\kappa < 0$, the nodes of $G$  
precede those of $F$ as $r$ increases. 
The same results can be obtained by considering 
the Ricatti equation for the ratio $F/G$ and its asymptotic values, as 
shown in \cite{rose1,rose2} for vector potentials.

As hinted in Eq.~(\ref{nodes}), the case when $n_F=0$ requires a special 
attention. Indeed, it can be shown that when the radial parts of the wave 
functions $G$ and $F$ have no nodes, the corresponding bound states can 
appear only in the $j=\ell+1/2$ state ($\kappa < 0$), but not in the 
$j=\ell-1/2$ state ($\kappa > 0$). This result was obtained previously 
in \cite{hirooka} for attractive scalar potentials of Coulomb type 
and separately for vector potentials of similar type. It is 
possible to generalize this result to the scalar and vector potentials 
under discussion. 
To prove the statement we note that according to Eq.~(9a), 
$GF$ is negative for $r$ sufficiently large. If both $G$ and $F$ have no 
nodes, then the relation $GF<0$ holds for any $r$. Since by Eqs.~(9b,c), 
$GF$ is negative near the origin only for states with $\kappa < 0$, 
it follows from continuity that bound states without nodes ($n_F=n_G=0$) 
must have $\kappa<0$, and hence appear only in 
the $j=\ell+1/2$ state, but not in the $j=\ell-1/2$ state. An example is 
shown in Fig.~1f for the case of the $0g_{9/2}$ $(\kappa=-5)$ intruder 
state.

We have obtained in Eq.~(\ref{condpos}) that the quantity $B(r)$ is positive 
at the nodes of $F$ and $G$. Even if the potentials $V_S$ and $V_V$ support 
only eigenstates with no nodes, there must still be a region where 
$B(r)>0$, in order that bound states exist. This can be inferred from 
the fact that as shown $GF$ is an increasing negative function at 
large $r$, and to enable it to vanish at $r=0$, its derivative 
$\left ( GF \right )'$ must change sign from positive to negative in 
some region. A glance at Eq.~(\ref{GFprime}) shows that since $A(r)>0$, 
a necessary (but not sufficient) condition for $\left ( GF \right )' $ 
to become negative is that 
\ba
B(r) > 0\;\; 
{\rm for\; some\; r}~.
\label{delta}
\ea
The above condition, $-\left [\,V_S(r)+V_V(r)\, \right ] > M-E$, means 
that in order that bound states exist, there has to be a region where the 
depth of the average attractive single-nucleon potential is larger than 
the binding energy.  

All of the above results are relevant for understanding properties of 
states in the relativistic pseudospin scheme \cite{gino,ami}. 
The generators for the relativistic pseudospin SU(2) algebra,
$\hat{\tilde{S}}_{\mu}$,
which commute with the
Dirac Hamiltonian, $[\,H\,,\, \hat{\tilde{S}}_{\mu}\,] = 0$, for the case when
$V_S= -V_V$ are given by \cite{bell}
\ba
{\hat{\tilde {S}}}_{\mu} =
\left (
\begin{array}{cc}
\hat {\tilde s}_{\mu} &  0 \\
0 & {\hat s}_{\mu}
\end{array}
\right )
= \left (
\begin{array}{cc}
U_p\, {\hat s}_{\mu} \, U_p & 0 \\
0 & {\hat s}_{\mu}
\end{array}
\right ) ~,
\label{Sgen}
\ea
where
${\hat s}_{\mu} = \sigma_{\mu}/2$ are the usual spin generators,
$\sigma_{\mu}$ the Pauli matrices, and
$U_p = \, {\mbox{\boldmath $\sigma\cdot p$} \over p}$ is the
momentum-helicity unitary operator introduced in \cite {draayer}. 
If in addition the potentials are spherically symmetric, 
the Dirac Hamiltonian has an additional invariant
SU(2) algebra, $[\,H\,,\, \hat{\tilde{L}}_{\mu}\,] = 0$, with the
relativistic pseudo-orbital angular momentum operators given by
\ba
\hat{\tilde{L}}_{\mu} =
\left (
\begin{array}{cc}
\hat {\tilde \ell}_{\mu} &  0 \\
0 & {\hat \ell}_{\mu}
\end{array}
\right )
= \left (
\begin{array}{cc}
U_p\, {\hat \ell}_{\mu} \, U_p & 0 \\
0 & {\hat \ell}_{\mu}
\end{array}
\right ) ~,
\label{Lgen}
\ea
where ${\hat\ell}_{\mu} = 
\mbox{ \boldmath $r$}\times \mbox{ \boldmath $p$}$ 
are the usual orbital angular momentum operators. 
The sets $\left\{\hat{\tilde S}_{\mu},\;
\hat{\tilde s}_{\mu},\;\hat{s}_{\mu}\right\}$ 
and $\left \{\hat{\tilde L}_{\mu},\; \hat{\tl}_{\mu},\; 
\hat{\ell}_{\mu}\right\}$ 
form two triads of $SU(2)$ algebras. 
The $\hat{\tilde S}_{\mu}$ and $\hat{\tilde L}_{\mu}$ operators 
act on the four-components Dirac wave functions. 
The ${\hat{\tilde s}}_{\mu}$ and ${\hat{\tilde l}}_{\mu}$ operators form 
the non-relativistic pseudospin and pseudo-orbital angular momentum 
algebras respectively, and act on the upper components of the 
Dirac wave functions. The ${\hat s}_{\mu}$ and ${\hat \ell}_{\mu}$ 
act on the ``small'' lower components of the Dirac wave functions. 
The total angular momentum operators 
${\hat J}_{\mu} = {\hat {\tilde L}}_{\mu} + {\hat{\tilde S}}_{\mu}$ 
have block diagonal form with entries 
${\hat j}_{\mu} = {\hat {\tilde \ell}_{\mu}} + {\hat {\tilde s}_{\mu}}
= U_p\,(\, {\hat \ell}_{\mu} + {\hat s}_{\mu} \, )\, U_p =
{\hat \ell}_{\mu} + {\hat s}_{\mu} $. 
The eigenfunctions of the Dirac Hamiltonian
are also eigenfunctions of 
$\mbox{\boldmath $\hat{\tilde{S}}\cdot \hat{\tilde{S}}$}$, 
$\mbox{\boldmath $\hat{\tilde{L}}\cdot \hat{\tilde{L}}$}$, 
and $\mbox{\boldmath ${\hat J}\cdot{\hat J}$}$, with eigenvalues 
${\tilde s}({\tilde s}+1)$, $\tl(\tl+1)$ and $j(j+1)$ respectively. 
The pseudospin ${\tilde s}=1/2$ and pseudo-orbital angular momentum $\tl$ 
are seen from Eqs.~(\ref{Sgen})-(\ref{Lgen}) to be the ordinary spin and 
ordinary orbital angular momentum of the lower component of the Dirac wave 
functions. This algebraic structure 
implies that the two states in the pseudospin doublet will have 
Dirac wave functions of the form 
\bmath
\ba
\Psi_{\kappa<0,m} &=& 
\left ({G_{\kappa<0}(r)\over r}
[Y_{\tl-1}\,\chi]^{(j)}_m\,,\,
i{F_{\kappa<0}(r)\over r}
[Y_{\tl}\,\chi]^{(j)}_m \right) \quad 
\;\;\;\; \kappa = -\tl < 0,\;\; j =\tl-{1\over 2}\quad 
\label{reldouba}\\
\Psi_{\kappa^{\prime}>0,m} &=& 
\left ({G_{\kappa^{\prime}>0}(r)\over r}
[Y_{\tl+1}\,\chi]^{(j^{\prime})}_m\,,\,
i{F_{\kappa^{\prime}>0}(r)\over r}
[Y_{\tl}\,\chi]^{(j^{\prime})}_m \right) \quad 
\kappa^{\prime} = \tl +1 > 0,\;\; j'=\tl+{1\over 2} ~.
\label{reldoub}
\ea
\emath
The members of the doublet share a common pseudo orbital angular momentum 
$\tl$, which is coupled to a pseudospin $\tilde{s}=1/2$. 
The state with negative $\kappa < 0$ involves unaligned pseudospin, 
$j=\tl - 1/2$, while its partner with positive $\kappa^{\prime} > 0$ 
involves aligned pseudospin, $j^{\prime}=\tl + 1/2$.

In the pseudospin symmetry limit the two states in Eq.~(15) form a 
degenerate doublet ($S=1/2$), and are connected by the pseudospin 
generators $\hat{\tilde S}_{\mu}$ of Eq.~(\ref{Sgen}). 
The corresponding upper components are a doublet 
with respect to the set $\hat{\tilde s}_{\mu}$ 
(the non-relativistic pseudospin algebra). Since the latter, 
by definition, intertwine space and spin, they can connect states for 
which the upper components have different radial wave functions, 
$G_{\kappa<0}(r)\neq G_{\kappa^{\prime}>0}(r)$. On the other hand, 
the corresponding lower components are a doublet with respect 
to the ordinary spin ${\hat s}_{\mu}$, and hence, in the pseudospin limit, 
their radial wave functions are equal up to a phase, 
\ba
F_{\kappa<0}(r) = F_{\kappa^{\prime}>0}(r) ~.
\label{twof}
\ea
In particular, $F_{\kappa<0}(r)$ and $F_{\kappa^{\prime}>0}(r)$ 
have the same number of nodes, which we denote by $n_r$. If we now use 
the result of Eq.~(\ref{nodes}), we find for $n_r\neq 0$ that 
$G_{\kappa<0}(r)$ in Eq.~(15) has also $n_r$ radial nodes, 
while $G_{\kappa^{\prime}>0}(r)$ has $n_r-1$ nodes. 
This explains the structure of nodes in the 
pseudospin doublets of Eq.~(\ref{psdoub}). 
The simple relation in Eq.~(\ref{twof}) between the radial wave functions 
of the lower components of the two states in the doublet, dictates 
this particular relation between the radial nodes of the corresponding  
upper components. This result cannot be obtained if one considers just 
the non-relativistic pseudospin algebra, $\hat{\tilde s}_{\mu}$, and is a 
direct outcome of the behavior of nodes of Dirac bound states 
and the identification of pseudospin as a relativistic symmetry 
of the Dirac Hamiltonian.

As we have shown, bound Dirac states, for which both the upper and 
lower components have no nodes ($n_r=0$) can occur only for 
$\kappa < 0$ and not for $\kappa^{\prime}> 0$. From 
Eq.~(\ref{reldouba}) we find that such states have pseudo-orbital 
angular momentum $\tl$ and total angular momentum $j = \tl -1/2$. 
As mentioned, these intruder states are ignored in the non-relativistic 
pseudospin scheme, and it is only the relativistic interpretation of
pseudospin symmetry, combined with known properties of Dirac bound 
states, which enable a classification for these states, as well as 
provide a natural explanation why these states do not have a 
pseudospin partner which is an eigenstate of the Hamiltonian.

The exact pseudospin limit requires that $V_S(r) = -V_V(r)$, which 
implies that $B(r) = E-M$ in Eq.~(3b). It is clear that under such 
circumstances the condition of Eq.~(\ref{delta}) cannot be fulfilled 
for bound states with positive binding energy $M-E>0$. This explains 
why in the exact pseudospin limit, there are no bound Dirac states 
and, therefore, by necessity the pseudospin symmetry must be broken 
in nuclei. Nevertheless, a variety of realistic mean field calculations
show that the required breaking of pseudospin symmetry in nuclei is 
small \cite{gino2,ring,arima,ginoami}. 
Quasi-degenerate doublets of normal-parity states and abnormal-parity 
levels without a partner eigenstate persist in the spectra, and the 
relation of Eq.~(\ref{twof}) is obeyed to a good approximation, 
especially for doublets near the Fermi surface. 
As discussed, these features are sufficient to ensure the observed 
structure of nodes occurring in pseudospin doublets 
and the special status of intruder levels in nuclei.

In summary, we have shown that identification of pseudospin as 
a relativistic symmetry of the Dirac Hamiltonian provides a natural 
explanation for the structure of radial nodes occurring 
in pseudospin doublets of normal-parity states. The key point in this 
explanation is that pseudospin symmetry implies a simple relation 
between the radial wave functions of the lower 
components of the two states in the doublet, and those in turn 
govern the radial nodes of the corresponding upper components. 
The intruder abnormal-parity states which have so far been discarded in 
non-relativistic treatments, can be accommodated in the relativistic 
pseudospin scheme and are assigned a pseudo-orbital quantum number 
$\tl$ and $j=\tl-1/2$. General properties of Dirac bound states provide a 
natural explanation for why these nodeless states do not have a partner 
which is an eigenstate of the Hamiltonian. 
It is gratifying to note that characteristic features 
({\it e.g.} radial and angular momentum quantum numbers) of states in the 
non-relativistic pseudospin scheme, which seem at first ad-hoc 
without an apparent reason, receive a proper justification once the 
relativistic origin of pseudospin symmetry in nuclei is taken into 
consideration.

We thank Dr. Jiri Mares for assistance with his numerical code. 
This research was supported in part by the U.S.-Israel Binational Science 
Foundation and in part by the United States Department of Energy under 
contract W-7405-ENG-36.

\begin{figure}
\noindent Figure 1. (a) The radial upper component ($G$) and lower component 
($F$) in (Fermi)$^{-1/2}$ and (b) the corresponding 
product $GF$ in (Fermi)$^{-1}$ 
of the $2p_{3/2}$ $(\kappa=-2)$ state. 
(c) and (d) The same for the $1f_{5/2}$ $(\kappa=3)$ state. 
(e) and (f) The same for the $0g_{9/2}$ $(\kappa=-5)$ state.
All states are eigenstates of a Dirac Hamiltonian with scalar ($S$) and 
vector ($V$) potentials: $V_{S,V}(r)= 
\alpha_{S,V}\left [1+ \exp({r-R\over a})\right]^{-1}$ 
with $\alpha_S = -358$, $\alpha_{V}= 292$ MeV, $R=7$ fm, $a=0.6$ fm.
\end{figure}

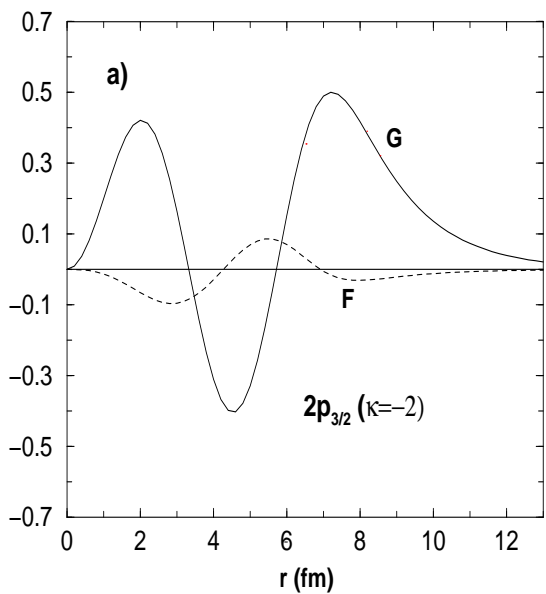
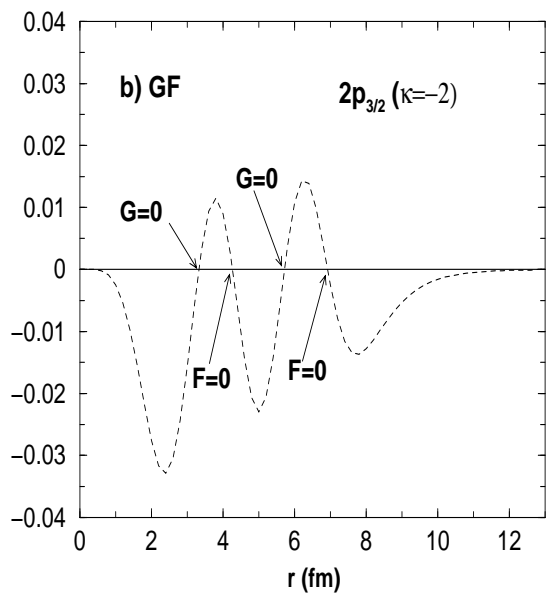
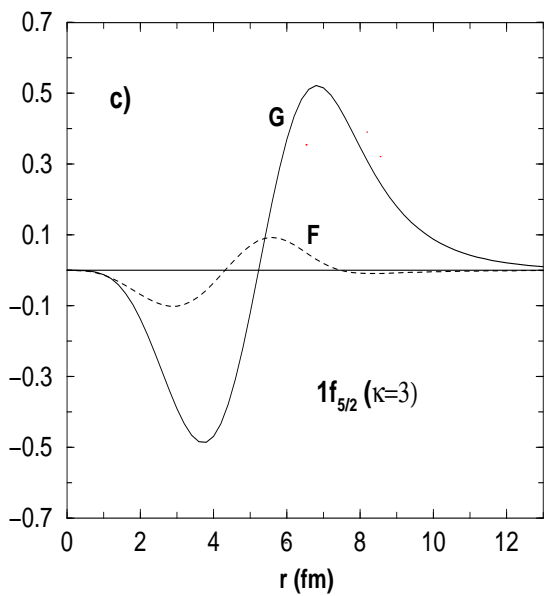
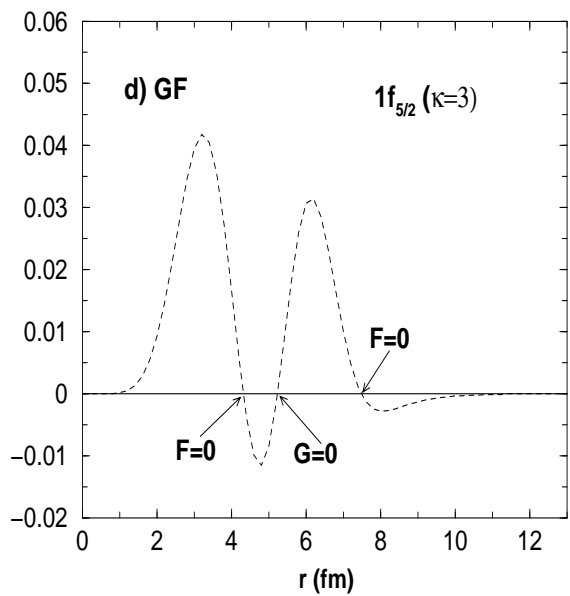
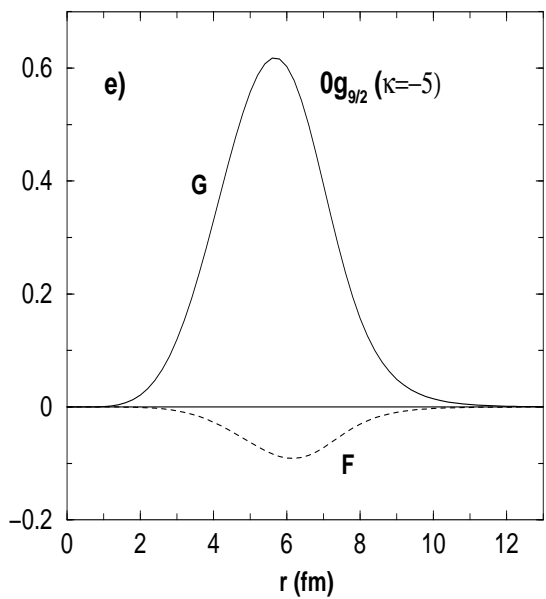
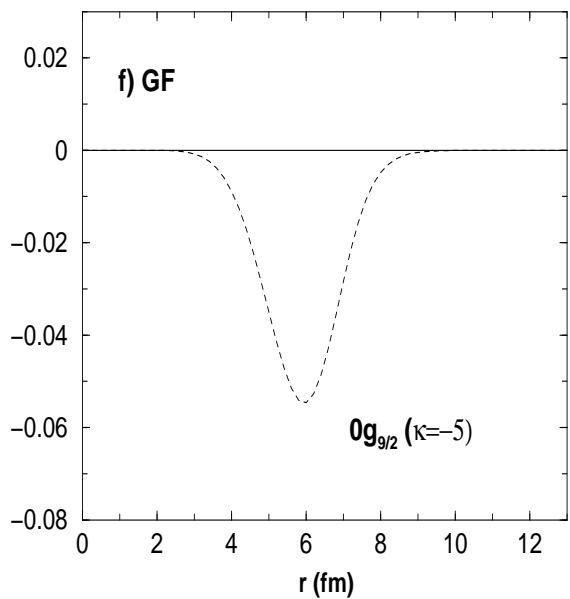
\end{document}